\documentclass[hyper,a4paper,12pt,oneside,notoc]{JHEP3}
\usepackage{amssymb,latexsym,amsmath}
\newcommand{\fracmy}[2] {\frac{#1}{#2}\:}
\newcommand{\sps} {\hspace{0.1cm}}
\newcommand{\dimi}[1] {~\hspace{-0.1cm} ^{(#1)} \hspace{-0.1cm}}
\newcommand{\sads}{\textit{AdSS$_5$}}
\newcommand{\adscft}{\textit{AdS/CFT}}
\title{AdSS$_5$ Brane World Cosmology}

\author{Denis Gon\c{t}a \\
Department of Physics, St. Petersburg State University, \\ St.
Petersburg, 198504, Russia \\ E-mail:
\email{d.gonta@pobox.spbu.ru}}

\abstract{The gravitational equations of the 5-dimensional
analogue of the \textit{AdSS} space-time, where all the matter
fields are confined on the 3-brane are examined. The most general
solutions are established in the generic case of a
non-$Z_2$-symmetric bulk. Constraining these solutions we derive a
number of remarkable metrics widely investigated in the
literature. Finally, we make many important conclusions about the
viability of the presented scenario and cosmology.}

\keywords{Brane Worlds, Schwarzchild-anti-de Sitter space-time}

\preprint{hep-ph/0401117}

\begin{document}

\section{Introduction}

Since the papers of Kaluza and Klein \cite{br1} it has been
suggested a possibility that there exist extra dimensions beyond
those of Minkowski space-time. In recent years, the ideas of extra
dimensions has become much more compelling. According to
\cite{br2,br3,br4} it has been suggested that additional
dimensions could have a quite distinct nature from those of the
Kaluza and Klein. In other words, ordinary matter would be
confined to our 4-dimensional universe while gravity would live in
extended space-time (bulk). During the past 30 years, research in
the theory of black holes has brought to light strong hints of a
fundamental relationship between gravitation, thermodynamics, and
the quantum theory. Indeed, the discovery of the thermodynamic
behavior of black holes has given rise to most of our present
physical insights into the nature of quantum phenomena occurring
in strong gravitational fields. Recently several authors have
found exact cosmological solutions of the brane worlds models
described by 5-dimensional black-hole like geometries
\cite{bh1,bh2,bh3,bh4,bh5,bh6,bh7,bh8,bh10}. Hence it is
significant to examine the bulk modifications in the case of the
non-$Z_2$-symmetry and the external finite temperature (see
below).

In a typical brane world scenario \cite{br2} our 4-dimensional
universe $q_{\mu \nu}$ is described by a 3-brane in 5-dimensional
space-time $g_{ab}$\footnote{Where the Latin indices $a,b =
0,1,2,3,4$; and the Greek indices $\mu,\nu = 0,1,2,3$.}. The extra
dimension need not be small or compact; in \cite{br3} Randall and
Sundrum (RS) shown that gravity can be localized on a single brane
even though the fifth dimension is noncompact. The metric contains
a \textit{warp factor} which is an exponential function of the
extra dimension
\begin{equation}
ds^2 = e^{-2k|y|}(-dt^2 + d\vec{x}^{\sps 2}) + dy^2.
\end{equation}
The bulk is described by the \textit{AdS$_5$} metric with $y = 0$
taken as the brane, so that $y < 0$ is identified with $y > 0$
illustrating thus the $Z_2$-symmetry.

The success of the {\adscft} correspondence
\cite{ads1,ads2,ads3,ads4,ads5} has lead to the intense study of
these space-times. The {\adscft} correspondence is a quite
specific conjecture. For instance, this conjecture allows us to
regard the entire $AdS_5$ geometry as a manifestation of the
dynamics of a 4-dimensional conformal field theory. It has been
realized that there are deep connections between the brane worlds
and the {\adscft} correspondence \cite{ads3,ads6,ads7}. In this
case, one should require two copies of the conformal field theory,
one for each of the $AdS_5$ patches. One important point is that
for a stable vacuum state of the theory, the extrinsic curvature
of the boundary surface (brane) should be proportional to the
induced metric \cite{ads2,ads4,ads8}. For this reason, the bulk is
changed from $AdS_5$ to {\sads} given by the metric
\begin{equation}\label{some-metric}
ds^2 = e^{-2y/\Lambda}[- \left(1- (\pi \Lambda T_0)^4
e^{4y/\Lambda} \right) dt^2+d\vec{x}^{\sps 2}]+ \frac{dy^2}{1-
(\pi \Lambda T_0)^4 e^{4y/\Lambda}},
\end{equation}
where $\Lambda \; (<0)$ is the bulk cosmological constant and
$T_0$ is the Hawking temperature (see \cite{ads2,bh0,bh1} for
discussions and references). This is a constant parameter of the
{\sads} solution. We stress that black-hole like geometries is an
standard way to introduce the temperature \cite{bh6,bh7}. At the
same time taking into account the brane worlds ideology one
suitable candidate can be constructed by gluing together two
patches of the {\sads} space-time along the brane world volume in
a $Z_2$-symmetric way \cite{bh1}
\begin{equation}\label{sads-metric}
ds^2 = e^{-2k|y|}[- \left( 1-\fracmy{U^4_T}{k^4}e^{4k|y|} \right)
dt^2+d\vec{x}^{\sps 2}]+
\frac{dy^2}{1-\fracmy{U^4_T}{k^4}e^{4k|y|}},
\end{equation}
where the horizon parameter $U_T$ is proportional to the external
temperature $T_0$.

Further, motivating by (\ref{some-metric}), (\ref{sads-metric})
and other possible \textit{AdSS$_5$} like space-times we assume
the ansatz
\begin{equation}\label{sads-anzatz}
ds^2 = e^{2\sigma(y)}[-\zeta(y)dt^2+d\vec{x}^{\sps 2}]+
\zeta(y)^{-1}dy^2.
\end{equation}
In fact, this anzatz gives a possibility to establish the most
general solution of such family of space-times. This solution can
be useful in order to clarify the symmetries and the main features
of the {\sads} bulk and therefore to derive the induced
Friedmann-Robertson-Walker (FRW) cosmology on the brane (see
\cite{ads2} for the radiation-dominated example). For the same
reason, it seems necessary to extend this scenario by adding
matter to the brane.

\section{Model Building}

Following methods of \cite{bh2,bh4} we suppose the location of the
brane in the form: $t=t(\tau)$, $y=y(\tau)$ parametrized by the
local time $\tau$ on the brane; then the induced metric on the
brane can be expressed as follows
\begin{equation}\label{ind0}
ds^2_{ind} = - d\tau^2 + a(\tau)^2 d\vec{x}^{\sps 2} \equiv q_{\mu
\nu} dx^\mu dx^\nu.
\end{equation}
Obviously, the embedding of a brane is specified by the extrinsic
curvature $K_{ab} = - (\delta^c_a - n_a n^c) \nabla_c \, n_b$ on
both sides, where $\nabla_*$ is the covariant derivative
compatible with (\ref{sads-anzatz}). Suppose the extrinsic
curvature being associated with the outward unit normal; then the
proportionality condition between the extrinsic curvature and the
induced metric takes the form \cite{ads2}
\begin{equation}\label{cond0}
K_{\mu \nu} = - \fracmy{1}{\Lambda} q_{\mu \nu}
\end{equation}
There is a set of solutions to (\ref{cond0}) that relates the
functions $t$ and $y$ in terms of $\sigma(y)$, $\zeta(y)$. In
other words, we have only one function $y(\tau)$ that locates the
brane. All the other surfaces have the same induced metric since
they are given by translations in $\tau$. Next step to proceed is
to parametrize a particular surface by $(t_*, \vec{x})$, where
$t_*$ is a solution of (\ref{cond0}) for $t$ in terms of $y$. Then
the induced metric becomes
\begin{equation}
ds^2_{ind} = -e^{2\sigma(y)}\zeta(y)dt_*^2 + e^{2\sigma(y)}
d\vec{x}^{\sps 2},
\end{equation}
which can be expressed in the standard FRW form (\ref{ind0})
defining the $y=y(\tau)$ function appropriately. Notice that the
authors of \cite{bh2,bh4} have introduced the relation between
$t(\tau)$ and $y(\tau)$ in the form of a handy normalization
condition. However in our case this relation is required by
(\ref{cond0}) in order to retain the vacuum state of the theory on
the boundary \cite{ads2,ads4,ads8}.

For simplicity, we consider the confinement of the matter on a
surface given by $\tau = \tau_*$ such that $y = y(\tau_*) = 0$,
but keeping in mind that all the other surfaces can be obtained by
translations in $\tau$. Then the 5-dimensional Einstein equations
are
\begin{eqnarray}\label{5d-einst-eqs}
\dimi{5}R_{ab}-\fracmy{1}{2}g_{ab} \dimi{5}R = \widetilde{k}^2
T_{ab}; \hspace{0.5cm} T_{ab} = -\Lambda g_{ab} +
\sqrt{|\zeta(y)|} \: S_{ab} \delta(y),
\end{eqnarray}
where $\widetilde{k}$ is the 5-dimensional gravitational coupling
constant such that $\widetilde{k}^2 = 8 \pi / \dimi{5}M_p^3$ with
$\dimi{5}M_p$ being the fundamental 5-dimensional Planck mass and
\begin{equation}\label{emt-decomposition}
S_{ab} = (- \lambda q_{\mu \nu} + \tau_{\mu \nu}) \delta^\mu_a
\delta^\nu_b.
\end{equation}
Notice that $\lambda \; (>0)$ and $\tau_{\mu \nu}$ are the
intrinsic brane tension and the energy-momentum tensor of the
brane that we re\-quire to be expressed in the perfect fluid form
$\tau_{\mu \nu} = diag\{ -\widetilde{\rho}, \widetilde{p_1},
\widetilde{p_2}, \widetilde{p_3}\: \}$. The equations
(\ref{5d-einst-eqs}) can be derived by taking the variation of the
action
\begin{equation}\label{action}
S = \int{d^{5}x \sqrt{|g|} \left( \fracmy{\dimi{5}M_p^3}{16\pi}
\dimi{5}R - \Lambda \right)} + \int{d^{4}x \sqrt{|q|} \left(
\left[ \fracmy{1}{2} \partial_\mu \varphi
\partial^\mu \varphi - V(\varphi) \right] - \lambda\right)}.
\end{equation}
The energy density and the pressure given by $\tau_{\mu \nu}$ are
expressed as follows
\begin{equation}
\widetilde{\rho} = -q_{00}\left( q^{00}\fracmy{1}{2}(\partial_0
\varphi)^2 - V(\varphi) \right) \equiv q_{00}\rho\: ,
\hspace{0.3cm} \widetilde{p}_\mu = -q_{\mu \mu}\left(
q^{00}\fracmy{1}{2}(\partial_0 \varphi)^2 + V(\varphi) \right)
\equiv q_{\mu \mu}p. \nonumber
\end{equation}
Suppose that $p = \omega \rho$; then it is straightforward to show
that the equations (\ref{5d-einst-eqs}) yield three relations
\begin{eqnarray}\label{einst-sol}
\zeta(y) \sigma'(y)^2 + \fracmy{1}{4}\zeta'(y)\sigma'(y) &=& k^2,
\notag \\ 2\zeta'(y)\sigma'(y) + \fracmy{1}{2}\zeta''(y) &=&
\widetilde{k}^2 \sqrt{|\zeta(y)|} (1 + \omega) \rho\: \delta(y),
\\ -3\sigma''(y)\sqrt{|\zeta(y)|} &=& \widetilde{k}^2 (\lambda + \rho)
\delta(y), \notag
\end{eqnarray}
where $k^2 \equiv -\widetilde{k}^2\Lambda/6$ and the prime denotes
differentiation with respect to $y$. We stress that these
relations are similar to the result for the first time obtained in
\cite{bh1} for $\rho = 0$.

It is possible to derive the most general solution of
(\ref{einst-sol}) in a closed form
\begin{eqnarray}\label{sol1}
\sigma(y) &=& \alpha |y| + \beta y + \chi; \label{sol1a} \\
\zeta(y) &=& \gamma^2 + \left( \fracmy{\widetilde{k}^2
\Lambda}{6(\alpha \epsilon(y) + \beta)^2} + \gamma^2 \right)
\left( e^{-4(\alpha |y| + \beta y)} - 1 \right) \label{sol1b}
\end{eqnarray}
with the Israel junction conditions \cite{br13} at the brane
\begin{eqnarray}\label{cond1}
\fracmy{\beta^2}{\alpha^2} &=& \fracmy{12 \Lambda}{\widetilde{k}^2
[2(\lambda + \rho)^2 - 3(\lambda + \rho)(1 + \omega)\rho]} + 1 ;
\quad \gamma^2 \alpha^2 = \frac{\widetilde{k}^4 (\lambda +
\rho)^2}{36}, \\ \text{where} &\quad& \lim_{y \to \pm 0} \zeta(y)
\equiv \gamma^2; \quad \lim_{y \to \pm 0} \sigma(y) \equiv \chi
\quad \text{and} \quad \epsilon(y) \equiv |y|\:' =
\begin{cases}
1,\; y>0 \\  -1,\; y<0
\end{cases}. \nonumber
\end{eqnarray}
It is seen that these solutions do not exhibit, in the general
case, the $Z_2$-symmetry and hence a number of restrictions should
be imposed. To be precise, suppose that $\alpha = -k, \: \beta =
0, \: \chi = 0$, and $\rho=0$; then (\ref{sol1a}), (\ref{sol1b})
and (\ref{cond1}) yield
\begin{equation}\label{sol2}
\sigma(y) = -k|y|,\quad \zeta(y) = 1,\quad \lambda = \left(
\frac{3\dimi{5}M^3_p}{4\pi} \right)k.
\end{equation}
This straightforwardly leads us to the warp factor and the
intrinsic brane tension exactly as in the original RS scenario
\cite{br2,br3}. It follows that in the $Z_2$-symmetric case with
no matter on the brane, we can achieve only an original RS like
scenario. If one relax the assumption of $Z_2$-symmetry ($\alpha =
-k, \: \beta \neq 0, \: \chi = 0, \: \rho=0$); then it can readily
be checked that (\ref{sol1a}), (\ref{sol1b}) and (\ref{cond1})
yield
\begin{eqnarray}\label{sol3}
\zeta(y) &=& \fracmy{k^2}{(\beta - k \epsilon(y))^2} - \left(
\fracmy{k^2}{(\beta - k \epsilon(y))^2} + \fracmy{k^2}{\beta^2 -
k^2} \right) e^{4(k |y| - \beta y)}, \label{sol3a}
\\ \sigma(y) &=& -k|y| + \beta y, \quad \text{where} \quad \beta^2
= k^2 \left(1 - \frac{9 \dimi{5}M_p^6 k^2}{16 \pi^2 \lambda^2}
\right). \label{sol3b}
\end{eqnarray}
Notice that one similar case was considered in the paper
\cite{bh1}. However the authors has focused on the equations like
(\ref{einst-sol}) being solved in the form of infinite series
\begin{eqnarray}\label{sol4}
\zeta(y) &=& 1 + \chi \zeta_1(y) + \chi^2 \zeta_2(y) + \chi^3
\zeta_3(y) + \chi^4 \zeta_4(y) + \mathcal{O}(\chi^5), \nonumber
\\ \sigma(y) &=& k|y| + \chi \sigma_1(y) + \chi^2
\sigma_2(y) + \chi^3 \sigma_3(y) + \chi^4 \sigma_4(y) +
\mathcal{O}(\chi^5) \nonumber
\end{eqnarray}
with $\zeta_i(y)$, $\sigma_i(y)$ illustrating both the
$Z_2$-symmetric and $Z_2$-antisymmetric behavior alternately.
Apparently, summing these infinite series one should obtain
expressions of the type (\ref{sol3a}), (\ref{sol3b}). The
inclusion of the matter gives us a possibility to consider the
case $\alpha = -k, \: \beta = 0, \: \chi = 0$, and $\rho \neq 0$.
The reader will have no difficulty in showing that the solution
(\ref{sol1a}), (\ref{sol1b}) can be reduced to the metric
(\ref{sads-metric}), where
\begin{equation}\label{sol5}
\sigma(y) = -k|y|, \quad \zeta(y) =
1-\fracmy{U^4_T}{k^4}e^{4k|y|};
\end{equation}
with two additional conditions in the form
\begin{eqnarray}\label{cond2}
\lambda &=& \fracmy{1}{4}\left( \rho(3\omega -1) \pm \sqrt{9
\rho^2 (1 + \omega)^2 - 12 \Lambda \dimi{5}M_p^3/\pi } \: \right),
\label{cond2a} \\ (1 + \omega) &=& \fracmy{1}{\rho} \left(
\fracmy{U_T^4 \dimi{5}M_p^3} {2 \pi k (k^4 + U_T^4)^\frac{1}{2}}
\right) \equiv \fracmy{1}{\rho} \theta > 0. \label{cond2b}
\end{eqnarray}
Now recall that the pressure is related to the cosmological energy
density via $\omega$. This implies that $p = - \rho + \theta$. In
fact, one can suppose that $\theta \ll \rho$; then $\rho \simeq -
p $ and the matter sector has a suitable form for the slow-roll
regime [see (\ref{action})].

Since we are interested in cosmological solutions, let us remember
that we considered the confinement of the matter on a surface
given by $\tau = \tau_*$ such that $y = y(\tau_*) = 0$. All the
other surfaces have the same induced metric (\ref{ind0}) related
by translations in $\tau$. The solutions (\ref{sol1a}),
(\ref{sol1b}) and the junction conditions (\ref{cond1}) are
relevant in order to understand symmetries and the main features
of the {\sads} bulk and therefore to derive the induced cosmology
on the brane. Typical brane worlds scenarios assume the $Z_2$
symmetry motivated by the M-Theory origin. Nevertheless many
recent papers examine models that are not derived from M-Theory.
For instance, there have been suggested multi-brane models and
scenarios, where the bulk is not manifestly $Z_2$-symmetric
\cite{bh2,bh4,bh5,bh10} or one basically different action from
(\ref{action}) is proposed \cite{bh11}. The conventional results
that underline the non-$Z_2$-symmetry reside in the effective
Friedmann equations. Notice that one general brane embedding
formalism have been constructed \cite{bh9} and the extra terms
beyond the standard $Z_2$-symmetric case, which characterize the
non-$Z_2$-symmetric embedding have been established. In our case,
we do not derive such field equations. However the proposed brane
model can be compared to the scenario \cite{bh2,bh4}, where one
similar space-time is taken. Then the extra term must behaves like
a positive Lambda-term for radiation or like a negative curvature
term for dust (see also \cite{bh5}). Further, I avoid presenting
details since the calculations are not original to me. We stress
that constraints on extra terms from nucleosynthesis must be taken
into account as well \cite{bh5,bh11}. It can be expected that
effects of the $Z_2$-symmetry breaking terms are decreasing with
the evolution of universe and hence at late times the standard
brane worlds cosmology should be recovered \cite{bh5,bh10}.

On the other hand, the bulk metrics (\ref{some-metric}) and
(\ref{sads-metric}) contain the terms that are proportional to the
Hawking temperature. This can be understood as an external
temperature that, in general, can be related to the boundary
conformal field temperature via the {\adscft} correspondence
\cite{ads2}. In our case, the expression (\ref{cond2b}) can
suggests one possible candidate of such relation. However the
finite temperature universe and the temperature induced on the
brane are the topics in the course of development
\cite{bh6,bh7,bh8} and hence we leave the details for a
forthcoming paper. Usually, the authors evaluate the energy of the
quantum bulk matter fields on a $AdS_5$(\sads) background at
nonzero temperature. The brane seems these thermal quantum effects
via the effective potential that supplements the effective
energy-momentum tensor.  Finally, the existence of the thermal
Casimir brane effect is also a subject of growing interest
\cite{bh6,ads4} and should be taken into account.

\section{Conclusions}

In this paper we have examined a brane world cosmological scenario
based on the {\sads} space-time. Taking account of the {\adscft}
correspondence and the black-hole analogy we noticed a number of
brane worlds candidates (\ref{some-metric}), (\ref{sads-metric})
(see also \cite{bh2,bh3,bh4,bh6}), which can be formulated in the
{\sads} bulk and further, we introduced the anzatz
(\ref{sads-anzatz}). In the basic section we have obtained the
most general solutions for this anzatz and the action
(\ref{action}). These solutions, in the general case, do not
exhibit the $Z_2$-symmetry and a number of restrictions should be
imposed. We have shown that the solutions (\ref{sol1a}),
(\ref{sol1b}) can lead us to the solutions of the original RS
scenario \cite{br2,br3} or to the setup of the scenario proposed
in \cite{bh1}.

The {\sads} brane cosmology is an widely investigated topic in the
literature. However in our case, the {\adscft} generic condition
(\ref{cond0}) imposes one physically relevant relation between the
location functions $t(\tau)$ and $y(\tau)$. Hence it can be
important to revise {\sads} based cosmologies taking account of
the solutions (\ref{sol1a}), (\ref{sol1b}) and the relation
(\ref{cond0}). Anyway, the arguments in this paper extend the
results in the literature on the black-hole like geometries and
the induced brane cosmologies.

\acknowledgments

The author is grateful to V.A. Franke and S.I. Vacaru for
collaboration and helpful discussions. The author thanks also E.V.
Prokhvatilov and S.A. Paston for support.


\begin{thebibliography}{25}

 \bibitem{br1}
  T. Kaluza
  {\it Sitz. Preuss. Akad. Wiss. Phys. Math.} K {\bf 1} (1921) 966;
  O. Klein {\it Nature} {\bf 118} (1926) 516

 \bibitem{br2}
  L. Randall and R. Sundrum
  {\it Phys. Rev. Lett.} {\bf 83} (1999) 3370 [hep-ph/9905221]

 \bibitem{br3}
  L. Randall and R. Sundrum
  {\it Phys. Rev. Lett.} {\bf 83} (1999) 4690 [hep-th/ 9906064]

 \bibitem{br4}
  T. Shiromizu, K. Maeda and M. Sasaki
  {\it Phys. Rev.} D {\bf 62} (2000) 024012 [gr-qc/9910076]

 \bibitem{br13}
  W. Israel
  {\it Nuovo Cimento} B {\bf 44} (1966) 4349



 \bibitem{bh0}
  S. W. Hawking and D. N. Page
  {\it Comm. Math. Phys.} {\bf 87} (1983) 577

 \bibitem{bh1}
  D. K. Park, H. Kim, Y. G. Miao and H. G. M\"{u}ller-Kirsten
  {\it Phys. Lett.} B {\bf 519} (2001) 159 [hep-th/0107156]

 \bibitem{bh2}
  P. Kraus
  {\it J. High Energy Phys.} {\bf 12} (1999) 11 [hep-th/9910149]

 \bibitem{bh3}
  P. Bowcock, C. Charmousis and R. Gregory
  {\it Class. Quant. Grav.} {\bf 17} (2000) 4745 [hep-th/0007177]

 \bibitem{bh4}
  D. Ida
  {\it J. High Energy Phys.} {\bf 09} (2000) 014 [gr-qc/9912002]

 \bibitem{bh5}
  A. C. Davis, I. Vernon, S. C. Davis, W. B. Perkins
  {\it  Phys. Lett.} B {\bf 504} (2001) 254 [hep-ph/0008132];
  S. C. Davis, W. B. Perkins, A. C. Davis, I. Vernon
  {\it Phys. Rev.} D {\bf 63} (2001) 083518 [hep-ph/0012223]

 \bibitem{bh6}
  S. Nojiri, S. D. Odintsov
  {\it Phys. Lett.} B {\bf 493} (2000) 153 [hep-th/0007205];
  S. Nojiri, S. D. Odintsov, S. Zerbini
  {\it Class. Quant. Grav.} {\bf 17} (2000) 4855 [hep-th/0006115]

 \bibitem{bh7}
  S. Shankaranarayanan
  {\it Phys. Rev.} D {\bf 67} (2003) 084026 [gr-qc/0301090]

 \bibitem{bh8}
  I. Brevik, K. A. Milton, S. Nojiri, S. D. Odintsov
  {\it Nucl. Phys.} B {\bf 599} (2001) 305 [hep-th/0010205]

 \bibitem{bh9}
  L. $\acute{A}$. Gergely
  {\it Phys. Rev.} D {\bf 68} (2003) 124011 [gr-qc/0308072]

 \bibitem{bh10}
  N. Deruelle, T. Dolezel
  {\it Phys. Rev.} D {\bf 62} (2000) 103502 [gr-qc/0004021]

 \bibitem{bh11}
  R. Cordero, A. Vilenkin
  {\it Phys. Rev.} D {\bf 65} (2002) 083519 [hep-th/0107175]



 \bibitem{ads1}
  O. Aharony, S.S. Gubser, J. Maldacena, H. Ooguri, Y. Oz
  {\it Phys. Rept.} {\bf 323} (2000) 183 [hep-th/9905111]

 \bibitem{ads2}
  S. S. Gubser
  {\it Phys. Rev.} D {\bf 63} (2001) 084017 [hep-th/9912001]

 \bibitem{ads3}
  M. J. Duff and J. T. Liu
  {\it Phys. Rev. Lett.} {\bf 85} (2000) 2052 [hep-th/0003237]

 \bibitem{ads4}
  V. Balasubramanian, P. Kraus
  {\it Commun. Math. Phys.} {\bf 208} (1999) 413 [hep-th/9902121]

 \bibitem{ads5}
  H .Verlinde
  {\it Nucl. Phys.} B {\bf 580} (2000) 264 [hep-th/9906182]

 \bibitem{ads6}
  L. Anchordoqui, C. Nunez, K. Olsen
  {\it J. High Energy Phys.} {\bf 10} (2000) 050 [hep-th/0007064]

 \bibitem{ads7}
  S. W. Hawking, T. Hertog, H. S. Reall
  {\it Phys. Rev.} D {\bf 62} (2000) 043501 [hep-th/0003052]

 \bibitem{ads8}
  N. S. Deger, A. Kaya
  {\it J. High Energy Phys.} {\bf 05} (2001) 030 [hep-th/0010141]




\end{thebibliography}
\end{document}